\documentclass[pra,twocolumn,showpacs,floatfix,superscriptaddress,aps]{revtex4}
\usepackage{amsmath}
\usepackage{makeidx}
\usepackage{amssymb}
\usepackage{graphicx}

\usepackage[usenames]{color}

\begin{document}

\title{Self trapping  of a dipolar Bose-Einstein condensate in a double well}


\author{S. K. Adhikari \footnote{adhikari@ift.unesp.br; URL  http://www.ift.unesp.br/users/adhikari}
}
\address{Instituto de F\'{\i}sica Te\'orica,
UNESP - Universidade Estadual Paulista,\\ 01.140-070 S\~ao Paulo, S\~ao
Paulo, Brazil}

\begin{abstract}

We study the Josephson oscillation and self trapping dynamics 
of a cigar-shaped 
dipolar Bose-Einstein condensate of 
$^{52}$Cr atoms polarized along the symmetry axis of an axially-symmetric  double-well potential using the numerical solution of a mean-field model, for dominating repulsive contact interaction (large positive scattering length $a$)
over an anisotropic dipolar interaction.  
Josephson-type oscillation emerges for small and very large number of atoms,
whereas self trapping is noted for an intermediate number of atoms. The dipolar interaction pushes the system away from  
self trapping towards  Josephson oscillation.
We consider a simple two-mode description for a qualitative understanding of the dynamics.

 \end{abstract}


\maketitle

\section{Introduction}

After 
the  observation of Bose-Einstein condensates (BEC) of 
$^{52}$Cr \cite{pfau,4}, $^{164}$Dy \cite{7}, and $^{168}$Er \cite{8}
atoms 
with large magnetic dipolar interaction,
there has been renewed activity   in the theoretical and experimental studies of 
degenerate gases.  The atomic interaction in usual nondipolar atoms is taken as an isotropic  $S$-wave contact interaction. The dipolar interaction, on the other hand, is  
anisotropic, long-range, and nonlocal acting in all partial waves. 
Due to the anisotropic nonlocal nature of  dipolar interaction, 
the stability of a
dipolar BEC 
depends on the number of atoms, the strength of dipolar interaction, the 
scattering length,  as well as, reasonably strongly and distinctly, on 
the trap geometry \cite{4,parker}.  
Among the
novel features noted in  a dipolar BEC, 
one can mention the peculiar  red-blood-cell-like
biconcave shape in density due to 
roton-like excitation \cite{roton} near the route to collapse, 
anisotropic sound and shock wave
propagation \cite{shock},
anisotropic D-wave collapse
\cite{collapse}, anisotropic soliton, vortex soliton \cite{soliton}
and vortex lattice formation \cite{lattice},  and anisotropic Landau critical velocity
\cite{landau} among others. Distinct stable checkerboard, stripe,
and star configurations in dipolar BECs have been identified in a two-dimensional (2D) optical lattice as stable Mott insulator \cite{mott} as well as superfluid soliton \cite{15}
states.

In a remarkable study, Smerzi {\it et al.} \cite{smerzi} predicted the dynamical trapping of a {\it repulsive} cigar-shaped BEC in one of the wells of a double-well potential using a simple two-mode description of the wave function 
for repulsive nonlinearity beyond a critical value and for an initial population imbalance between the two wells.  This is counter-intuitive as a repulsive BEC is expected to expand and occupy both the wells equally.  The phenomenon of self trapping, appearing  due to the self-interaction of a BEC, and of Josephson oscillation
have been studied extensively in nondipolar BECs \cite{aubry,exptst}. There have also been    studies of Josephson oscillation and 
self trapping of a dipolar BEC in a double well \cite{xiong}  and in a toroidal trap \cite{abad}.

In this paper we study self trapping in a cigar-shaped dipolar BEC in a double well, aligned along the polarization direction, in the presence of an anisotropic nonlocal dipolar and a repulsive contact interaction. 
The BEC is subject to a strong radial and weak axial trap and we use a reduced one-dimensional (1D) model for the description of its dynamics \cite{luca,SS,dell,lp}.
We consider a large enough value of scattering length $a$, so that the net self-interaction in the dipolar BEC is repulsive. The interplay between the repulsive contact interaction and the anisotropic  long-range dipolar interaction will make this study more challenging and of general interest. Although, the dipolar interaction is directional, possibly because of maintaining an axial symmetry for  easy theoretical analysis,  in most experiments on dipolar atoms \cite{pfau,4,7}, 
the polarization direction is taken along the axial symmetry direction. For most theoretical investigations, this symmetric set-up has been used and the role of self-interaction in dipolar BEC is well understood in this configuration. For this reason we shall consider the polarization direction along the axial symmetry direction in this study. Although, we shall not  study self trapping in a fully anisotropic dipolar BEC, we shall make qualitative remarks about expected results in such cases.

In the thoroughly studied dipolar BEC of $^{52}$Cr atoms,  the strength of the repulsive contact interaction is stronger than the dipolar interaction so that the net interaction is repulsive and is suitable for the study of self trapping. The same is not obvious in a strongly dipolar $^{164}$Dy BEC, where the net interaction could be attractive to make the system unsuitable for self trapping. This is why we consider the interaction parameters of $^{52}$Cr atoms in this study. Moreover, it is possible to manipulate the scattering length to a lower value by the Feshbach resonance technique \cite{fesh}, to make the dipolar BEC barely repulsive for studying the interplay of contact and dipolar interactions in self trapping.

In this study of self trapping of a cigar-shaped repulsive dipolar BEC in a double-well potential, the following general trend is established. The total number of atoms $N$ and the initial population imbalance $S_0\equiv (N_{10}-N_{20})/N$ play decisive 
roles in self trapping, where $N_{10}$ and $N_{20}$ are the initial number of atoms in the two wells: $N=(N_{10}+N_{20})$.  For an initial population imbalance $S_0$ greater than a critical value $S_c$ there is self trapping for $N$ beyond a critical number $N_c$ consistent with the prediction of the two-mode description \cite{smerzi}.  For $S_0<S_c$, there is  Josephson oscillation for all $N$. For $S_0>S_c$, there is  Josephson oscillation for $N<N_c$ and self trapping for $N_u>N>N_c$ where $N_u$ is an upper limit of $N$ for self trapping. 
The   Josephson oscillation is driven by the net repulsive nonlinear interaction, which increases with $N$ and facilitates  Josephson oscillation resulting in an increase of the  frequency of  oscillation 
with $N$.

In Sec. \ref{IIA} we present the time-dependent 3D mean-field model for the cigar-shaped dipolar BEC and in Sec. \ref{IIB} we present an effective 1D model for the same. In Sec. \ref{IIC} we present the two-mode description of the dynamics, which proves to be very useful for a qualitative understanding of the relevant features of self trapping and Josephson oscillation in a  dipolar BEC trapped in a double-well potential. 
The results of numerical simulation are presented in Sec. \ref{III} using the effective 1D model. Some of the results of the 1D model are confirmed by a numerical simulation of the full three-dimensional (3D) model. Finally, in Sec. 
\ref{IV} we present a summary and conclusion of the study.

\section{Analytical Consideration}
\label{II}

\subsection{3D mean-field model}

\label{IIA}
 
A dipolar BEC of $N$ atoms, 
each of mass $m$ satisfies the   mean-field
Gross-Pitaevskii (GP) equation \cite{pfau}
 \begin{eqnarray}  \label{gp3db} 
&& \mathrm{i}\hbar\frac{\partial}{\partial t} {\Phi{(\bf r},t)}
 =  \biggr[ -\frac{\hbar^2\nabla^2 }{2m} + V({\bf r}) 
 +\frac{4\pi \hbar^2 a}{m} |\Phi({\bf r},t)|^2
\nonumber \\
&&+ \frac{\mu_0 \mu_d^2}{4 \pi} \int U_{\mathrm{\mathrm {dd}}}({\bf R})|\Phi({\bf r'},t)|^2d{\bf r'}
\biggr] {\Phi({\bf r},t)}, 
\\
&& V({\bf r})=  \frac{m\omega_z^2 }{2}\left[ \frac{  \rho^2}{\lambda^2}
 + z^2  +2 A e^{-\kappa z^2}\right], 
\end{eqnarray} 
where $ {\bf R = (r-r')}, $
 $\Phi({\bf r},t)$ the wave function, 
  $a$  the atomic scattering length, $n({\bf r},t)
\equiv |\Phi({\bf r,}t)|^2$ the BEC density normalized as 
$\int n({\bf r},t) d{\bf r}=N$, 
 where 
$ \omega_\rho$ 
and $ \omega_z$ are the angular frequencies of radial and axial traps with aspect ratio $\lambda \equiv \omega_z /\omega_\rho $, 
 $\mu_0$ is the permeability  of free space and $\mu_d$ is the 
magnetic dipole moment of each atom. 
The constants $A$ and $\kappa$ are the strength and width of a 
Gaussian barrier responsible for the double-well potential.
The   dipolar interaction between two atoms at $\bf r$ and $\bf r'$ 
in Eq.  (\ref{gp3db}) 
is taken as 
\begin{equation}\label{xx}
 U_{\mathrm{\mathrm {dd}}}({\bf R}) = 
 \frac{(1-3\cos^2	\theta)}{R^3} , 
\end{equation}
 where $	\theta$ is the angle between the vector  $\bf R$ and the polarization 
direction $z$ taken along the axial symmetry direction.  
  
To compare the strengths of atomic short-range
and dipolar interactions, the dipolar interaction is often expressed in terms of the 
length scale $a_{\mathrm{\mathrm {dd}}}=m\mu_0\mu_d^2/(12\pi \hbar^2)$. Using this length scale, it is convenient 
to write the dipolar GP  equation (\ref{gp3db}) in the following dimensionless form
  \begin{align}  \label{3d1b} 
&  \mathrm{i}\frac{\partial}{\partial t} 
{\Phi{(\bf r},t)}
=  \biggr[ -\frac{\nabla^2 }{2} + V({\bf r})+4\pi a |\Phi({\bf r},t)|^2\nonumber \\
&
 +  3a_{\mathrm{\mathrm {dd}}}\int 
U_{\mathrm{\mathrm {dd}}}({\bf r -r'})
|\Phi({\bf r'},t)|^2d{\bf r'}
\biggr] {\Phi ({\bf r},t)}\\
& V({\bf r})=\frac{1}{2}
\left(\frac{\rho^2}{\lambda^2}+ z^2\right)+ Ae^{-\kappa z^2}. \label{tr}
\end{align} 
  In Eq.  (\ref{3d1b}) 
energy, length, density $n({\bf r})$ and time $t$ are expressed in units of 
oscillator energy
$\hbar \omega_z$, oscillator length 
$l_0\equiv \sqrt{\hbar/m \omega_z}$,  $l_0^{-3}$, and $t_0=\omega_z^{-1}$, respectively.

\subsection{1D reduction of the 3D mean-field model}

\label{IIB}

For  a cigar-shaped BEC with a strong radial and weak axial confinement, it is convenient to consider simplified equations in 1D for a description of the axial dynamics \cite{luca}.
For a 
{dipolar BEC} with a
strong radial trap ($\lambda^2 << 1$), we assume that in the
radial direction the BEC is confined in the ground state
$\Psi({\bf \rho}) = \exp[-\rho^2/(2\lambda)]/\sqrt {\lambda \pi}$ 
 of the transverse trap and the wave function   $\Phi({\bf r},t)
 = \Psi_{1D} (z,t)
\Psi({\bf \rho})$ can be written as \cite{SS,dell,deu,lp}
\begin{eqnarray}\label{anz1d}
\Phi({\bf r},t)&=& \frac{1}{\sqrt{\pi \lambda}}\exp \left[-\frac{\rho ^2}
{2\lambda}\right] \Psi_{1D}(z,t).
\end{eqnarray}
The interesting relevant axial dynamics is carried by the wave function 
$\Psi_{1D}(z,t)$.
The density 
in configuration space $n({\bf r},t)\equiv |\Phi({\bf r},t)|^2  $ is related to that in momentum space $\widetilde n({\bf k},t)$ 
by  the Fourier transformation
\begin{eqnarray}
\widetilde n({\bf k},t)&\equiv &\int e^{{\mathrm i}{\bf k}\cdot {\bf r}} |\Phi({\bf r},t)|^2
d{\bf r}= 
 \widetilde n({\bf k_\rho})\widetilde n({ k_z},t),\\
\widetilde n({ k_z},t)&\equiv &\int e^{{\mathrm i}{ k_z} {z}} |\Psi_{1D}(z,t)|^2 dz,\\
\widetilde n({\bf k_\rho})&=&\int e^{{\mathrm i}{\bf 
k_\rho}\cdot {\bf \rho}} |\Psi({\bf \rho})|^2
d{\bf \rho}\equiv  \exp\left[- \frac{k_\rho^2 \lambda }{4} 
   \right].
\end{eqnarray}
The Fourier transformation and its inverse are  defined by
\begin{align}
&\widetilde A({\bf k})=\int d{\bf r}A({\bf r})  e^{\mathrm{i}{\bf k}\cdot {\bf r}},\quad
 A({\bf r})=\int \frac{d{\bf k}}{(2\pi)^3}\widetilde A({\bf k})  e^{-\mathrm{i}{\bf k}\cdot {\bf r}}.
\end{align}

To derive the effective
1D equation for the cigar-shaped {dipolar BEC}, we substitute the
ansatz (\ref{anz1d}) in Eq.  (\ref{3d1b}), multiply by the ground-state wave
function $\Psi(\rho)$ and integrate over $\rho$ to get
the 1D equation \cite{lp}
\begin{align}\label{gp1d}
& \mathrm{i}\frac{\partial \Psi_{1D}(z,t)}{\partial
t}=\biggr[-\frac{\partial_z^2}{2}+
V(z)+ 
\frac{2 a}
{\lambda 
}\vert\Psi_{1D}(z,t)\vert^2
\nonumber \\ &+ \frac{2a_{\mathrm {dd}}}{\lambda }
\int_{-\infty}^{\infty}\frac{dk_z}{2\pi}e^{ik_z z}\widetilde
n(k_z)s\biggr(k_z \frac{\sqrt{\lambda}}{\sqrt 2}\biggr)\biggr]\Psi_{1D}(z,t) , \\
&\equiv  \biggr[-\frac{\partial_z^2}{2}
+V(z)+
 \frac{2 a \vert\Psi_{1D}(z,t)\vert^2}
{\lambda } \nonumber \\
& + \int_{-\infty}^{\infty} U_{\mathrm {dd}}^{1D}( Z )\vert\Psi_{1D}({z'},t)\vert^2d{z'}
\biggr] \Psi_{1D}(z,t), 
\end{align} 
with normalization $\int dz |\Psi_{1D}(z,t)|^2 =N$, where 
\begin{eqnarray}
V(z)= \frac{1}{2} z^2+Ae^{-\kappa z^2},  \label{sw}
\\
s(\zeta) = \int_0^\infty du \left[ \frac{3\zeta^2}{u+\zeta^2}-1
\right] e^{-u}.
\end{eqnarray}
The 1D potential in momentum and 
configuration spaces, $V_{1D}(k_z)$ and $U_{\mathrm{dd}}^{1D}(Z)$,  are, respectively, \cite{lp} 
\begin{align}  
V_{1D}(k_z)&=2 a_{\mathrm{dd}}\int_0^\infty dk_\rho k_\rho \left[
\frac{3k_z^2}{k_\rho^2+k_z^2}-1\right]
\exp\left[-\frac{k_\rho^2d_\rho^2}{2} \right],
\\& \equiv \frac{2a_{\mathrm{dd}}}{d_\rho^2}
s_{1D}\left(   \frac{k_zd_\rho}{\sqrt 2}\right),
\\& s_{1D}(\zeta)=\int_0^\infty du \left[ \frac{3\zeta^2}{u+\zeta^2}
-1\right]\exp(-u),
\\
 U_{\mathrm {dd}}^{1D}(Z) &= \frac{1}{2\pi}\int_{-\infty}^{\infty} dk_z e^{ik_z z} V_{1D}(k_z) \notag \\
& = \frac{6 a_{\mathrm {dd}}}{( {2\lambda}
)^{3/2}} \big[\frac{4}{3}\delta(\sqrt t) +2\sqrt t
-\sqrt
\pi (1+2 t) e^t {\mbox{erfc}}(\sqrt t)\big], \label{1dpotx}  
\end{align}
where $t=Z^2/(2 \lambda), Z=\vert z-z'\vert$ and where erfc is the complementary error function.
Similar, but not identical, 1D reduced potential was derived in \cite{SS,deu}, where the
$\delta$-function term in the 1D potential (\ref{1dpotx})
was not explicitly specified. However, this term is included in Ref. \cite{deu2}. 
Another distinct formulation of   1D reduction
of the dipolar GP equation is available \cite{bao1d}.

\subsection{Two-mode description of dynamics}
\label{IIC}

The main features of the dynamical evolution of the cigar-shaped 
 dipolar BEC in a double-well trap 
can be 
obtained by considering the following two-mode wave function 
\cite{smerzi}
\begin{equation}\label{2m}
\Phi({\bf r},t)=\sum _{i=1}^2\psi_i(t)\phi_i({\bf r}),
\end{equation}
where the normalizable function $\phi_i({\bf r})$ is strongly 
localized 
in well $i=1,2$ with uniform amplitude
$\psi_i(t)=\sqrt{N_i} e^{\mathrm{i}\theta_i(t)}$, where $N_i$ and $\theta_i$ 
are the number of atoms in the two wells and their respective phases.
Here we are using the 3D GP Eq. (\ref{3d1b}) in the two-mode description. An equivalent two-mode description can be formulated using the 1D GP Eq. (\ref{gp1d}).  
The condition of strong localization of the wave functions  $\phi_i({\bf r})$ implies
\begin{eqnarray} \label{x1}
\int \phi_i({\bf r})\phi_j({\bf r}) d{\bf r}=\delta_{ij}\\
\int f({\bf r}) \phi_1({\bf r})\phi_2({\bf r}) d{\bf r}=0,
\label{x2}
\end{eqnarray}
for any $f({\bf r})$.
This leads to the conservation of the number of atoms $N=(N_1+N_2)$.

Substituting Eq. (\ref{2m}) in Eq. (\ref{3d1b}) we obtain
\begin{align} \label{over}
&\mathrm{i}\sum_{i=1}^2 \dot \psi_i(t)\phi_i({\bf r})
=\sum_{i=1}^2 \left[- \frac{1}{2}\psi_i(t)\nabla^2 \phi_i({\bf r})
+   \psi_i(t)V({\bf r})\phi_i({\bf r})
 ] 
\right]\nonumber \\
&+ \sum_{i,j=1}^2\left[4\pi a  \phi_i^2({\bf r}) +3a_{\mathrm {dd}}\int U_{\mathrm {dd}}({\bf R}) \phi_i^2({\bf r}')d {\bf r}'\right] N_i \psi_j \phi_j({\bf r}),
\end{align}
where according to the strong localization conditions (\ref{x1}) and 
(\ref{x2}), we have neglected  the overlap integrals of the localized wave functions $\phi_i({\bf r})$. 

Multiplying Eq. (\ref{over}) by $\phi_j({\bf r})$, integrating 
over $\bf r$, and using the strong localization conditions (\ref{x1}) and 
(\ref{x2}), we get \cite{footnote}
\begin{align}
\mathrm{i}\dot \psi_i(t)=& [E_i+ A_{i1}N_1+A_{i2}N_2]\psi_i(t)-K\psi_j(t),\quad j\ne i,   
\label{20}\\
 \label{aij}
A_{ij}=&  4\pi a \delta_{ij}\int \phi_i^4 
({\bf r}) d{\bf r}\nonumber \\
+&
3a_{\mathrm {dd}}\int  d{\bf r }\int  d{\bf r '}
U_{\mathrm {dd}}({\bf R}) 
\phi_i ^2({\bf r}')\phi_j^2 ({\bf r}),\\
K=&-\int\left[ \frac{1}{2}\nabla \phi_1({\bf r})
 \nabla \phi_2({\bf r})+
\phi_1({\bf r}) \phi_2({\bf r}) V({\bf r})
\right]d{\bf r},\\
E_i=& \int \left[  \frac{1}{2}  (\nabla \phi_i({\bf r}))^2+
 \phi_i^2({\bf r}) V({\bf r})
\right]d{\bf r}.
\end{align}

In terms of the phase difference $\delta(t)=\theta_2(t)-\theta_1(t)$ and population imbalance $S(t)=[N_1(t)-N_2(t)]/N$, Eqs. 
 (\ref{20}) can be written as \cite{smerzi}
\begin{align}\label{eq1}
\dot S(t)&=-\sqrt{1-S(t)^2} \sin \delta(t),\\
\dot \delta(t) &= \Lambda S(t) +   \frac{S(t)}{\sqrt{1-S(t)^2}}\cos \delta(t)     + \Delta E,  \label{eq2}
\end{align}
where time has been rescaled as $2Kt\to t$ and where 
\begin{align}
\Delta E= \frac{E_1-E_2}{2K}+\frac{(A_{11}-A_{22})N}{4K},\\
\Lambda =\frac{(A_{11}+A_{22}-A_{12}-A_{21})N}{4K}. 
\label{lambda}
\end{align}

Equations (\ref{eq1})  and (\ref{eq2}) describe the oscillatory motion of the dipolar system and are quite similar to the same  for a nondipolar BEC, although in the present dipolar system there are contributions from the dipolar energy in the parameters  $A_{ij}$, viz. Eq. (\ref{aij}).  Equations (\ref{eq1})  and (\ref{eq2}) are to be solved from the initial condition: $S(0)=S_0, \delta(0)=\delta_0$.
Oscillatory motion through the value $S(t) =0$ is possible for 
small values of the parameter $\Lambda$. The oscillatory motion through the point $S(t)=0$ is stopped for \cite{smerzi}
\begin{eqnarray}\label{crit}
\Lambda> \Lambda_c \equiv 2 \frac{\sqrt{1-S(0)^2}\cos \delta(0)+1}
{S(0)^2}. 
\end{eqnarray}
The pendulum-like free oscillation of the atoms between the two wells is possible for $\Lambda < \Lambda_c$.

Equation (\ref{crit}) is fundamental in explaining qualitatively the onset of self trapping and also the role of dipolar interaction on it.  The constant $\Lambda_c$ reduces with the increase of $S(0)$. 
Hence self trapping is more likely for a large $S(0)$ and should disappear for $S(0)\to 0$. Also, from Eq. 
(\ref{lambda}) we see that $\Lambda \to 0$ as $N\to 0$. Hence, self trapping can only appear for the number of atoms $N$ larger than a critical value. To study the role of dipolar interaction on self trapping, we note that the constant $A_{ij}$ of Eq. (\ref{aij}) is two times the interaction energy of the system. The off-diagonal contribution to dipolar energy ($i\ne j$) is expected to be much smaller than the diagonal contribution ($i=j$) and hence can be neglected for a qualitative understanding of the dynamics. 
Here we are considering a cigar-shaped dipolar BEC, where the dipolar interaction energy given by the double integral in Eq. (\ref{aij})
is negative (attractive) and will reduce the values of the constants $A_{ii}$ and consequently the value of the constant $\Lambda $ given by Eq. (\ref{lambda}). With this reduction of the value of  $\Lambda $, the dipolar interaction will push the system away from self trapping 
as with a smaller $\Lambda$  it will be more difficult to satisfy  condition (\ref{crit}): $\Lambda > \Lambda_c.$  For very large number of atoms $\Lambda$ becomes very large and the condition (\ref{crit}) is always satisfied implying self trapping. Nevertheless, for a very large nonlinearity, the two-mode description breaks down 
 { even for a nondipolar BEC and its prediction becomes unreliable.   In this case, the repulsive nonlinear (contact) interaction
increases the chemical potential above the height of the inter-well
barrier. Consequently, the contribution of the double well in Eq. (\ref{3d1b}) can be neglected in comparison to that of  the nonlinear term and the loss of self trapping is expected in the absence of an effective double-well trap.}  We shall demonstrate 
these aspects of dynamics from the numerical solution of the mean-field model.

\section{Numerical Result}


\label{III}

With the above insight to oscillation dynamics from the two-mode description, we solve the mean-field model equation for the same. 
The results from the two-mode description is most reliable for small 
to medium values of contact and dipolar interaction energies. This is also the domain of validity of the 1D mean-field model, as was thoroughly established previously \cite{lp}
for statics and dynamics of a dipolar BEC. The full 3D mean-field model calculation of dynamics is prohibitively time consuming, hence in this study we use mostly the 1D mean-field model (\ref{gp1d}) to study the oscillation dynamics.  In certain cases we also solve the 3D GP Eq.
(\ref{3d1b}) and compare the results for dynamics with the results obtained from the 1D model.

We solve the GP equations  (\ref{3d1b}) or (\ref{gp1d}) numerically by the split-step Crank-Nicolson method \cite{CPC,12}.  { The dipolar integral is evaluated in the Fourier momentum
$({\bf k})$ space using convolution  as \cite{12} 
\begin{align} \label{con}
&\int d{\bf r}'V_{\mathrm {dd}}({\bf r-r}')n({\bf r}')
=\int \frac{d{\bf k}}{(2\pi)^3}e^{-\mathrm{i}{\bf k}\cdot {\bf r}}
\widetilde V_{\mathrm {dd}}({\bf k})\widetilde n({\bf k}) 
,
\end{align}
  The FT $\widetilde V_{\mathrm {dd}}({\bf k})$ of the dipole potential $ V_{\mathrm {dd}}({\bf r-r'})$ is analytically known in 3D
\cite{12} and numerically evaluated in 1D. The FT of density $n({\bf r})\equiv
|\Phi({\bf r})|^2$ is evaluated
numerically by means of a standard fast FT (FFT)
algorithm. The dipolar integral in Eq.   (\ref{3d1b}) or (\ref{gp1d}) 
 is
evaluated by the convolution  (\ref{con}). The inverse
FT is taken by the standard FFT algorithm.  
}
We use typically a space step of 0.1 and time step 0.001 in 3D and 
of 0.025 and 0.0005 in 1D and consider up to 
512 space points in  3D and 4096 in 1D for  discretization.

Before we present the results of self trapping for a dipolar $^{52}$Cr  BEC in a double-well potential, it is pertinent to describe the phenomenon of self trapping as previously 
considered in Ref. \cite{smerzi} as well as, for a dipolar BEC, 
  in Ref. \cite{xiong}. The self trapping is the surprising dynamical locking of a weakly-repulsive BEC 
in one of the wells of a double well \cite{smerzi}, while it is expected that such a 
BEC will occupy both the wells due to atomic repulsion. We emphasize that the self-trapped state is not an eigenstate of the   time-independent mean-field equation. 
It is natural that a weakly-attractive BEC can be locked 
in one of the wells 
due to atomic attraction corresponding to an eigenstate  of the   time-independent mean-field equation. 

The authors of Ref. \cite{xiong}  call the   stationary states of a cigar-shaped   attractive  dipolar BEC   localized in one of the wells of a double well as self-trapped states.  Such states are quite
different from the dynamically trapped nonstationary 
states of Ref. \cite{smerzi} in a repulsive cigar-shaped BEC. The authors of Ref. \cite{xiong}  suggest to vary the angle $ \varphi$ between the polarization direction and the axial $z$ axis in a cigar-shaped dipolar BEC with zero contact interaction.  For $\varphi=0$, the cigar-shaped BEC is oriented along the polarization direction, thus resulting in an attractive system.    For $\varphi=\pi/2$, the cigar-shaped BEC is oriented perpendicular to the polarization direction, thus resulting in a repulsive  system.  With the increase of the angle $\varphi$ from 0 to $\pi/2$, the system gradually becomes repulsive from attractive. For small $\varphi$, spontaneously symmetry broken stationary states localized in one of the wells of the double well
appear due to atomic attraction.  For large $\varphi,$ the system is repulsive thus leading to 
symmetric stationary states occupying both wells. This passage of symmetric to symmetry-broken states is termed self trapping in 
Ref. \cite{xiong}. The self-trapped states of the present paper are dynamically trapped states in a repulsive BEC in a double well as in Ref. \cite{smerzi} and {\it not} the stationary states of an attractive dipolar BEC
bound in one of the wells as in Ref. \cite{xiong}.

 To study dynamical self trapping in a repulsive cigar-shaped BEC, the  
symmetry-broken initial stationary state is taken as   that in the asymmetric double well  \cite{fermist}
\begin{equation}\label{asw}
V'({\bf r})= \frac{1}{2}(z-z_0)^2 +Ae^{-\kappa z^2} ,
\end{equation}
in place of (\ref{sw}).
The asymmetric 
ground (stationary) state in this asymmetric well is obtained
by solving the corresponding GP equation (\ref{gp1d}) by imaginary time
evolution. 
The parameter $z_0$ in Eq. (\ref{asw}) is chosen so that the population imbalance  $S(0)$ has a fixed predetermined value. 
We will study self trapping in the symmetric double well (\ref{sw})  
using the GP equation (\ref{gp1d}) with identical parameters used in generating the initial asymmetric stationary state. In actual experiment \cite{exptst}, the symmetry-broken initial state was prepared in this fashion.

In the present study, as in most experiments on dipolar atoms
 \cite{pfau,4,7}, we consider  $\varphi=0$ and $a>a_{\mathrm {dd}}$.  The dipolar length $a_{\mathrm {dd}}$ denotes the strength of dipolar interaction as the scattering length $a$ $ (>0)$ denotes the strength of the repulsive contact interaction. The condition    $a>a_{\mathrm {dd}}$ guarantees that the cigar-shaped dipolar BEC aligned along the 
polarization direction $z$
is always repulsive and there cannot be any symmetry-broken stationary state.
This condition is satisfied for the dipolar atoms so far used in BEC experiments \cite{pfau,4,7,8}. 
We consider a $^{52}$Cr BEC with dipole moment 
$\mu_d=6 \mu_B$, with $\mu_B$ the Bohr magneton, so that the dipolar strength $a_{\mathrm {dd}}= 15a_0$ \cite{pfau}, with $a_0$  the Bohr radius. In our calculation we take the oscillator length 
$l_0 =1$ $\mu$m corresponding to  the axial angular frequency $\omega_z \approx 2\pi \times 194$ Hz. To generate a cigar-shaped dipolar BEC \cite{1d}
in the double well we take the parameter $\lambda = 1/9$
corresponding to an 
angular frequency of the transverse radial trap   $\omega_\rho= 2\pi \times 1746$ Hz. 
The parameters of the double well (\ref{tr}) are taken as $A=16, $ and $\kappa=
10$. The width and height of the Gaussian barrier in the double well has to be appropriate for allowing a smooth Josephson oscillation. A very wide and a very high barrier will  substantially hinder the Josephson oscillation and facilitate self trapping. On the other hand, a very narrow  and a very low barrier will tend to reduce the double well to a single well and, hence, should hinder self trapping and facilitate Josephson oscillation.  Otherwise, these parameters ($A$ and $\kappa$) do not have much influence on the phenomenon of self trapping and Josephson oscillation of a repulsive dipolar BEC in a double well.  These values of the parameters $A$ and $\kappa$ 
of the double well were used previously for a satisfactory study of self trapping in a Fermi superfluid at unitarity \cite{fermist}.

For $^{52}$Cr atoms $a_{\mathrm{\mathrm {dd}}}=15a_0$ and to maintain the 
net interaction in the cigar-shaped dipolar BEC repulsive for $\varphi =0$, 
 we shall consider two values of the scattering length $a$ $(>a_{\mathrm{\mathrm {dd}}})$: $a=20a_0$ and $100a_0$. The scattering length can be manipulated in laboratory by varying a background magnetic field near a Feshbach resonance \cite{fesh}.
There are three domains of the initial population imbalance  $S(0)$ in the double well  which we consider in the following:

\noindent{(a) {\it small} $S(0)$:
The numerical calculations  show that 
there is no self trapping for a small $S(0)$ $ (<0.1)$ in the dipolar BEC. The  two-mode description  
(\ref{crit}) reveals that  a small $S(0)$ leads to a large $\Lambda_c$,  which can be attained for a large $N$ for a fixed $a$ and $a_{\mathrm{\mathrm {dd}}}$, viz. Eq. (\ref{lambda}). { Nevertheless, for a large $N$, the nonlinear interaction energies in the GP equation (\ref{3d1b}) become large and the  role of the Gaussian barrier in this equation becomes very small and can be neglected. Consequently, the double well essentially reduces to a single well allowing for smooth pendulum-like oscillation. In this limit of small $S(0)$  and large $N$
the two-mode description breaks down.  }
}

\noindent{
(b) {\it medium} $S(0)$:
  For a slightly larger $S(0)$  $ (0.15 \lesssim S(0) \lesssim  0.25)$,
prediction (\ref{crit}) leads to a small to moderate 
$\Lambda_c$, which can be attained for a medium value of $N$  in a 
nondipolar BEC ($a_{\mathrm {dd}} =0$)
within the validity of the two-mode description.  Consequently, there is self trapping in a nondipolar BEC
as will be
confirmed in the numerical calculation. For these intermediate values of $S(0)$, the attractive dipolar interaction tends to cancel the repulsive contact interaction and stops the constant $\Lambda$ of Eq. (\ref{lambda}) attain the critical value 
$\Lambda_c$ of Eq. (\ref{crit}) except for very large $N$ leading to 
large dipolar and contact nonlinear interactions, when the two-mode description becomes unreliable.  Consequently, for  $0.15\lesssim S(0) \lesssim0.25$    there  is no self trapping in the dipolar system for $a=20a_0, a_{\mathrm {dd}}=15a_0$, whereas self trapping appears in the nondipolar system with 
$a=20a_0, a_{\mathrm {dd}}=0 $ as reported below.} 

\noindent{(c) {\it large} $S(0)$:
For larger $S(0)$ $(0.3\lesssim S(0)\lesssim 1)$, the critical value    $\Lambda_c$ of Eq. (\ref{crit}) is small and there is self trapping in all cases: dipolar or nondipolar.}

\begin{figure}[!t]
\begin{center}

\includegraphics[width=\linewidth]{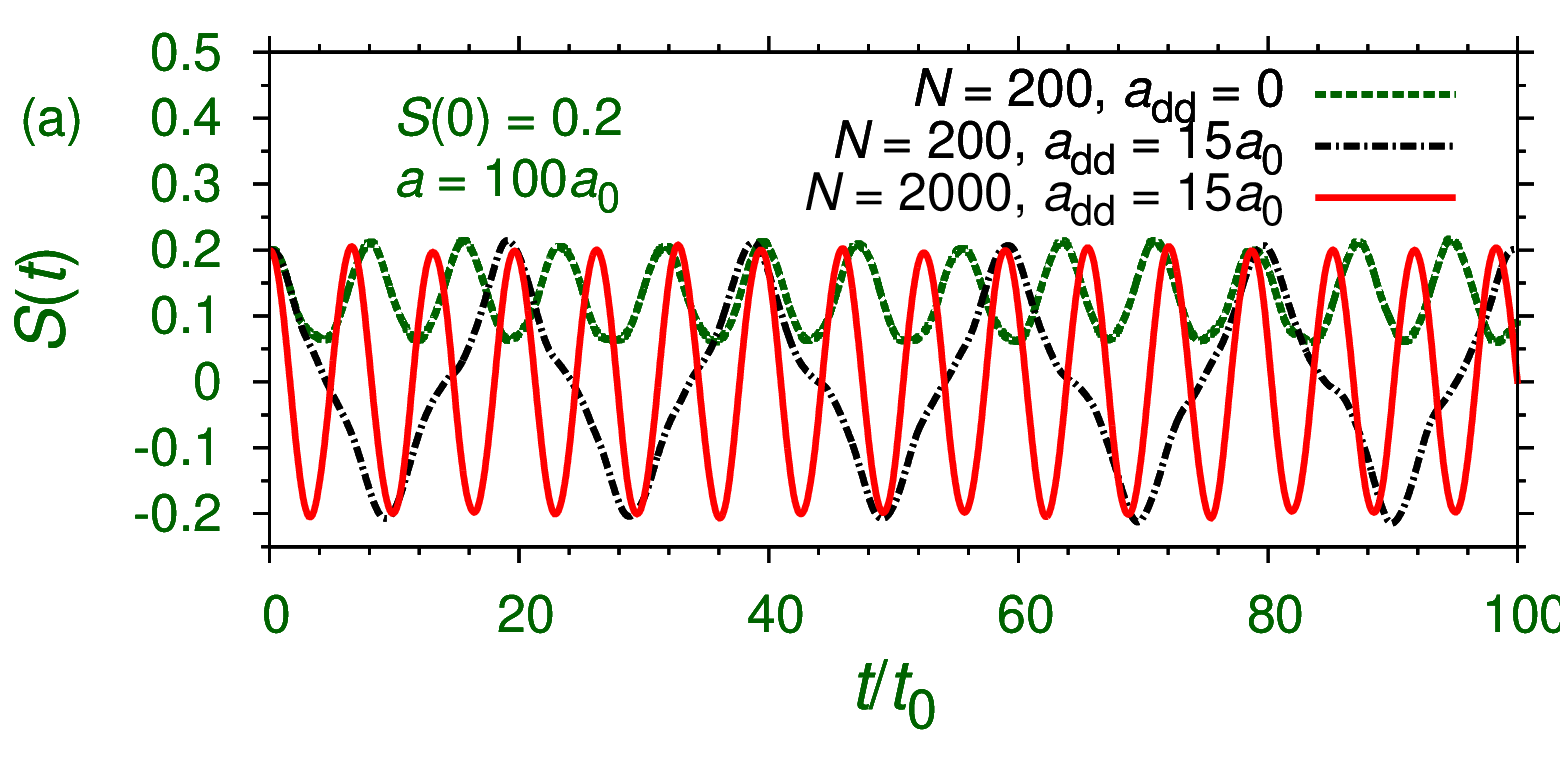}
\includegraphics[width=\linewidth]{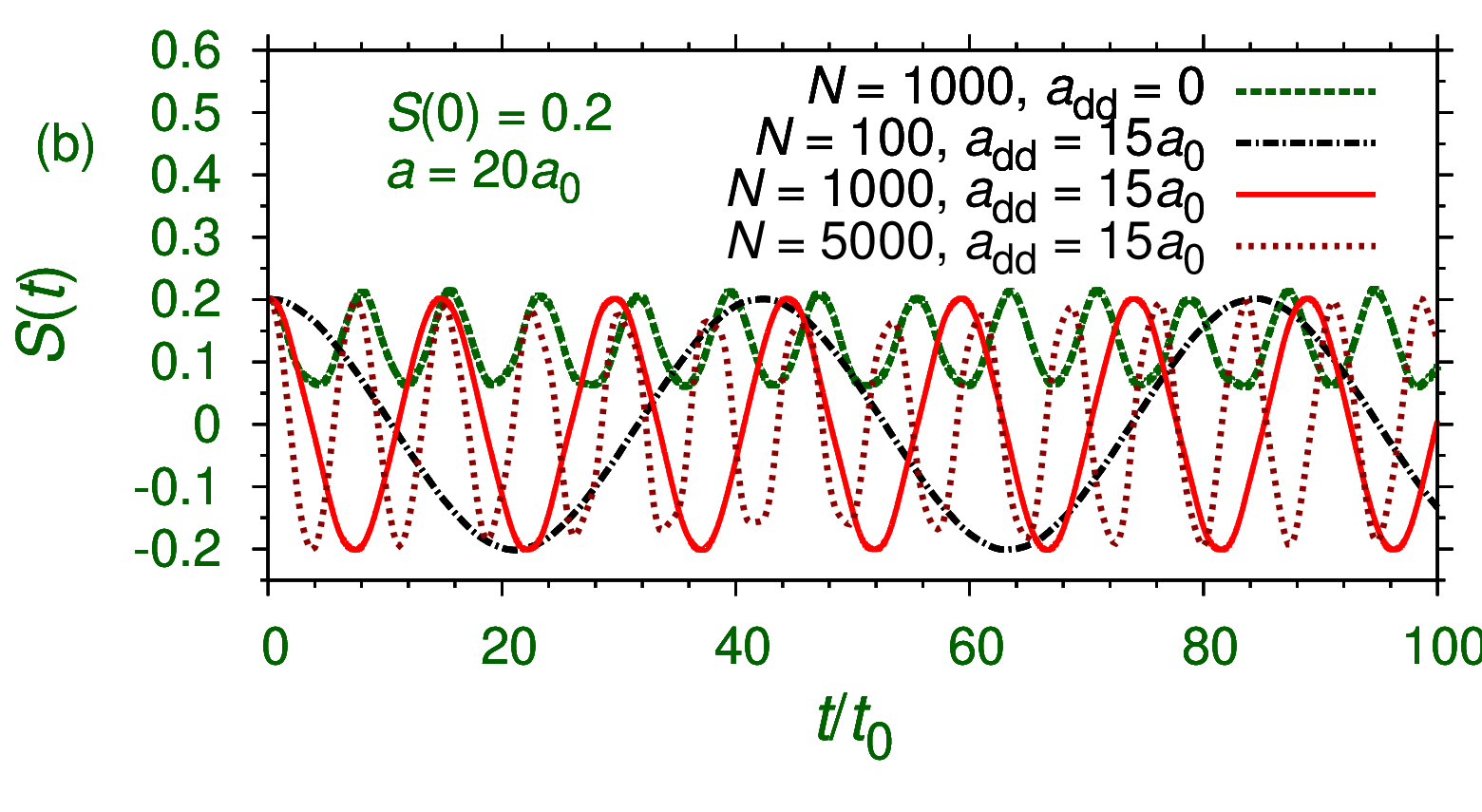}
\includegraphics[width=\linewidth]{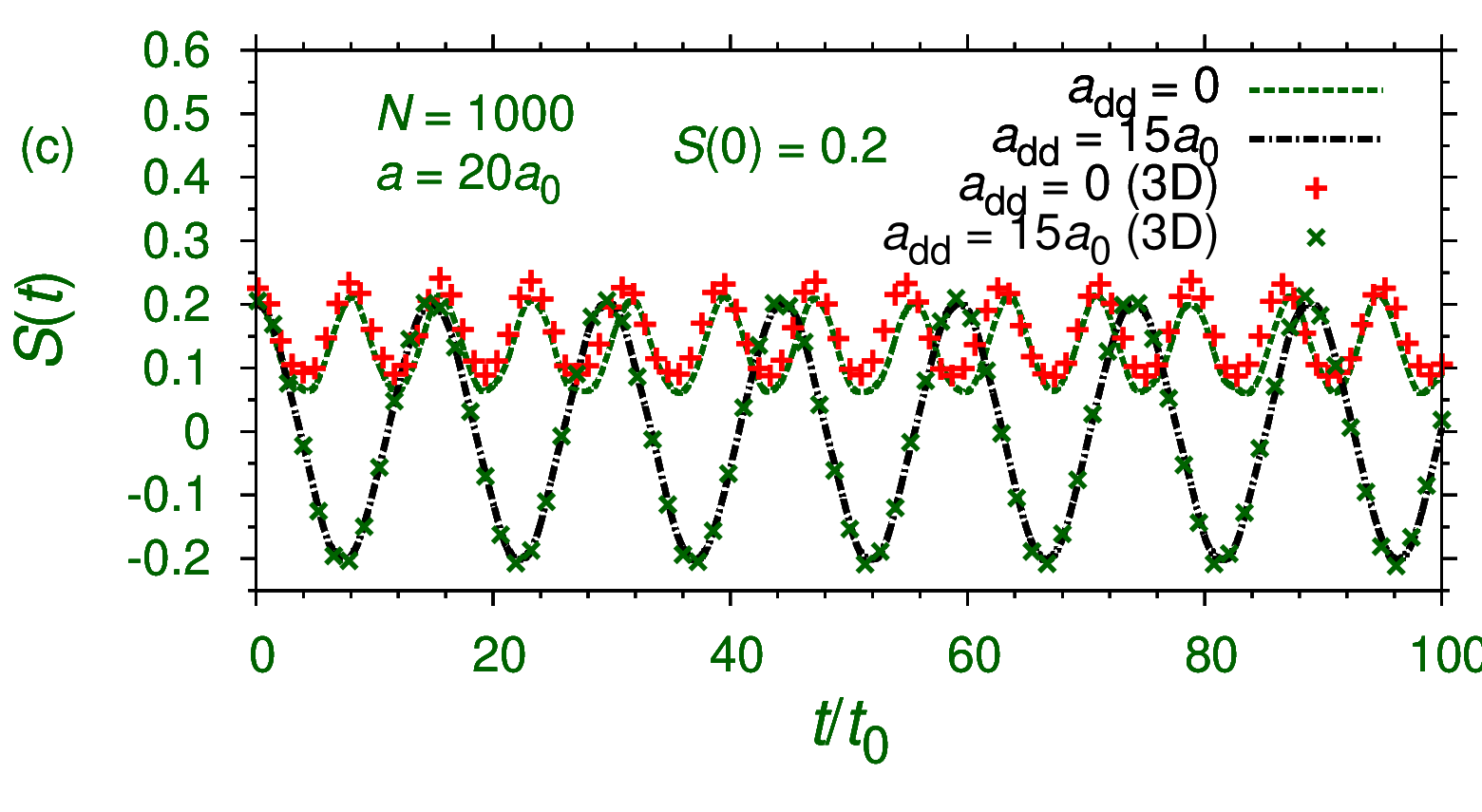}

\caption{ (Color online)  Population imbalance $S(t)$ versus time $t/t_0$ 
for $S(0)=0.2$ and  (a) $a=100a_0$ and (b) $a=20a_0$,  
 for different $N$ and $a_{\mathrm {dd}} =0$ and $15a_0$ as obtained from a numerical simulation of the 1D model (\ref{gp1d}). (c)  Population imbalance $S(t)$ versus time $t/t_0$ for $N=1000, a=20a_0, S(0)=0.2$ and $a_{\mathrm{dd}}=0, 15a_0$ from the 1D (full line) and 3D
(chain of symbols) models.
  }\label{fig1}
\end{center}
\end{figure}

An initial state with the desired initial population imbalance 
$S(0)$ is prepared by solving the 1D GP equation (\ref{gp1d}) by imaginary-time propagation
with the asymmetric well (\ref{asw}) in place of the symmetric well 
(\ref{sw}). With this initial state we perform the real-time propagation of the 1D GP equation (\ref{gp1d}) with the symmetric well (\ref{sw}) maintaining all other parameters (dipolar and nondipolar interactions and the number of atoms) unchanged throughout the numerical simulation.
In laboratory this is achieved by preparing a BEC in the asymmetric well and then 
suddenly changing the trapping potential from asymmetric to symmetric and observing the subsequent dynamical evolution. The self trapping and Josephson oscillation is best illustrated in a dynamical evolution of the population imbalance $S(t)$.
In Figs. \ref{fig1} (a) and (b), we plot $S(t)$ versus $t/t_0$ for $S(0)=0.2$ and for $a=100a_0$ and $20a_0$, respectively.  In both cases  self trapping is possible in the nondipolar system ($a_{\mathrm {dd}}=0$) resulting in  a positive time-averaged population imbalance $\langle S(t) \rangle$. However,  the dipolar $^{52}$Cr BEC ($a_{\mathrm {dd}}=15a_0$), 
permanently stays in the Josephson oscillation regime resulting in a null value of $\langle S(t) \rangle$. As Josephson oscillation is driven by the repulsive nonlinear interaction, an increase of the number of atoms 
corresponding to  a larger nonlinear interaction leads to a larger frequency as can be established in Fig. \ref{fig1} (a), comparing the results of $N=200$ and 2000 in the dipolar case for $a=100a_0$.  This is also evident 
in Fig. \ref{fig1} (b),   comparing the results of $N=100, 1000$ and 5000 in the dipolar case for $a=20a_0$.

\begin{figure}[!t]
\begin{center}\includegraphics[width=\linewidth]{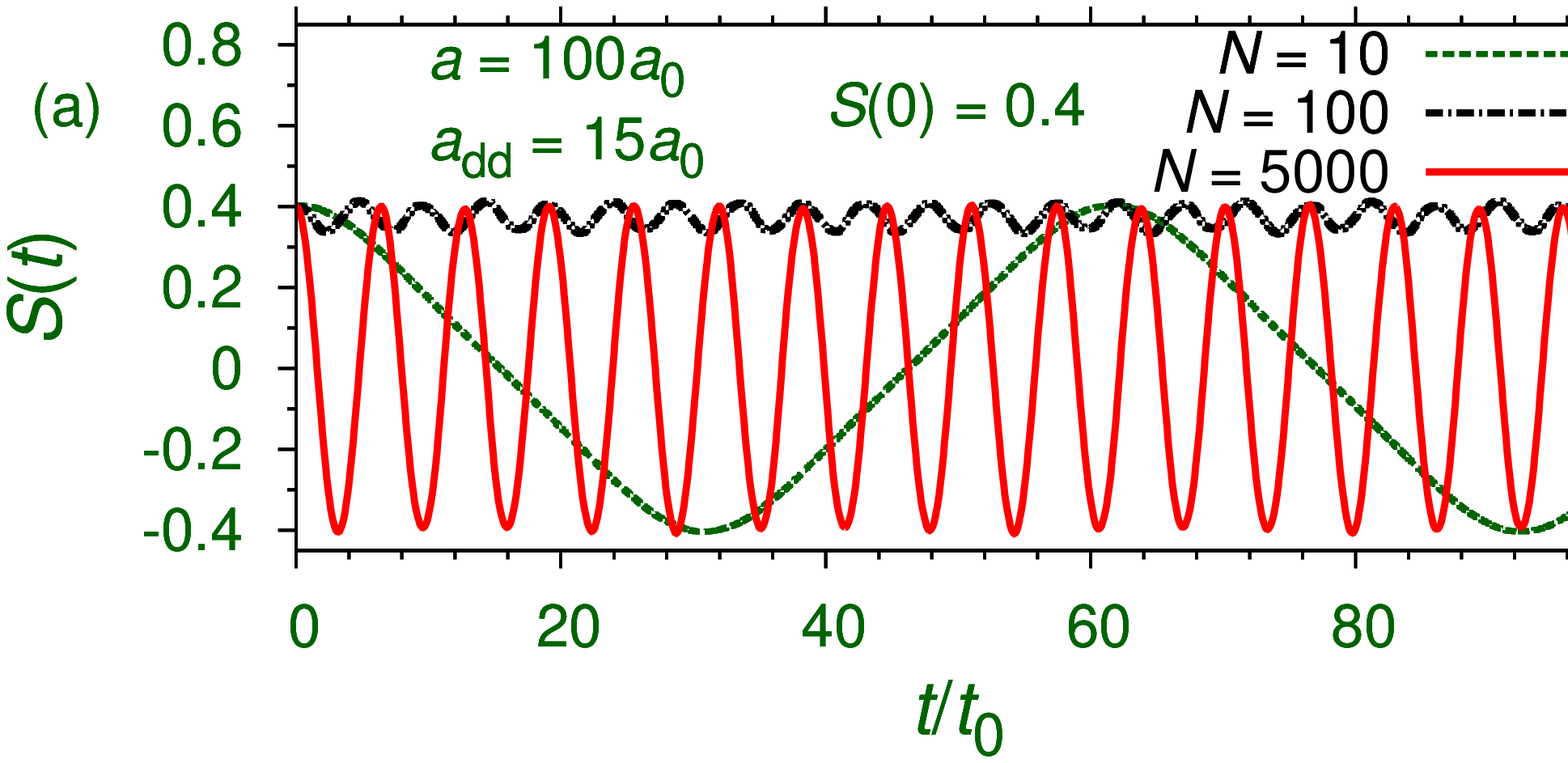}
\includegraphics[width=\linewidth]{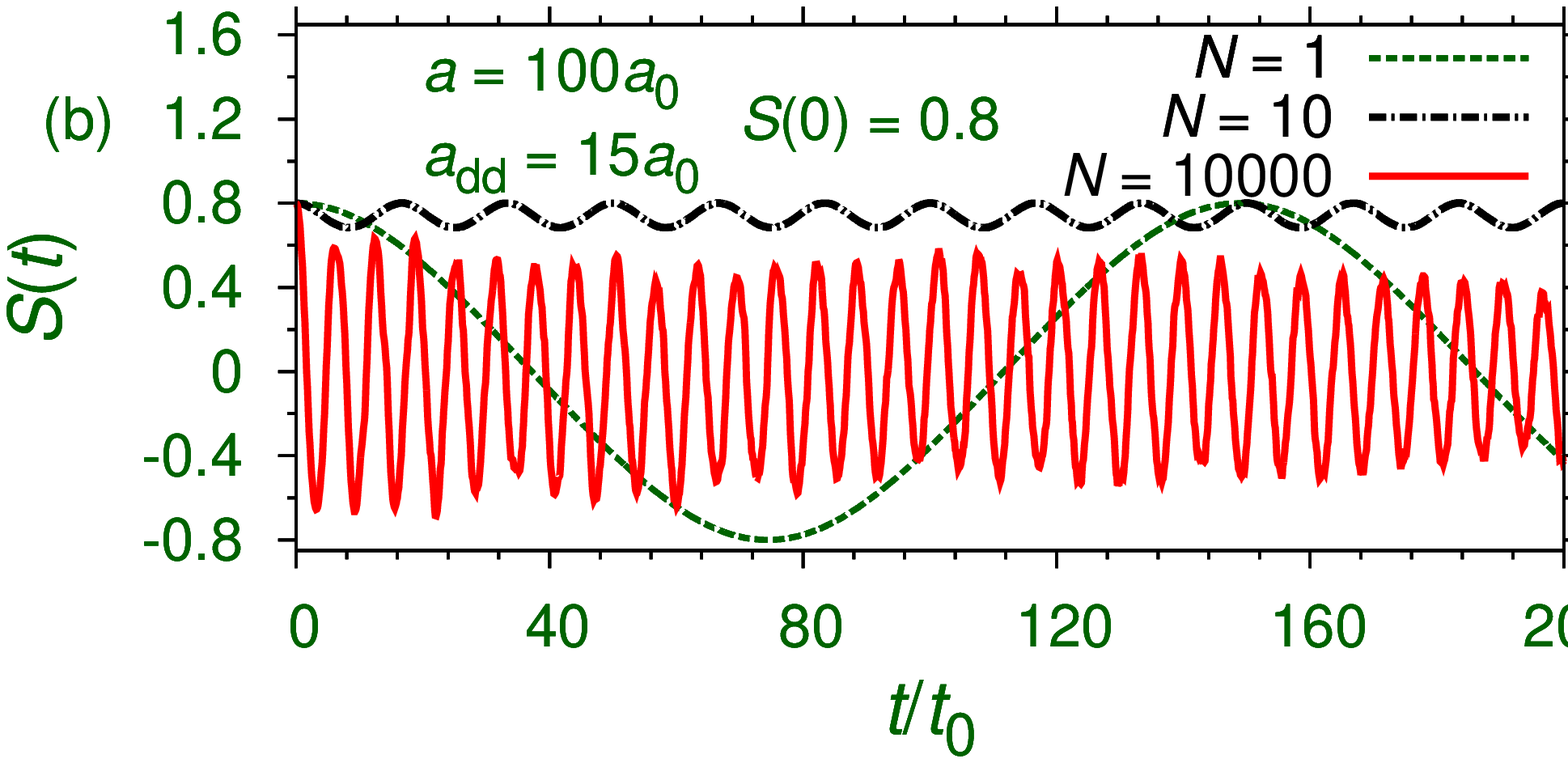}

\caption{ (Color online)  Population imbalance $S(t)$ versus time $t/t_0$ 
for  $a=100a_0, a_{\mathrm {dd}}=15a_0$ and for  (a)
 $S(0)=0.4$ and (b)  $S(0)=0.8$ and  
 for different $N$ using the 1D model (\ref{gp1d}).
  }\label{fig2}
\end{center}
\end{figure}

Next to see the reliability of the 1D description of the dynamics, we also solved the 3D Eq. (\ref{3d1b}) with potential (\ref{tr}) for $\lambda = 1/9$ and all other parameters remaining unchanged from the 1D model simulation, e. g., $N=1000, a=20a_0, S(0)=0.2, a_{\mathrm{dd}}=0,$ and $ 15a_0$.   The initial state in 3D was prepared with the following asymmetric well:
\begin{align}
V({\bf r})= \frac{1}{2}\left[ \frac{\rho^2}{\lambda^2}+(z-z_0)^2 \right]+ Ae^{-\kappa z^2},
\end{align}
in place of the trap (\ref{tr}). In the preparation of the initial state we use the same parameters as used in Eq. (\ref{asw}) in 1D.
The results for self-trapping and oscillatory dynamics as obtained from 1D and 3D simulations are compared in Fig. \ref{fig1} (c). 
The good agreement between the two simulations assures the reliability of the 1D model calculation.

Now  we consider a larger initial population imbalance $S(0)$. 
We present results   of self trapping and Josephson oscillation
of a cigar-shaped dipolar $^{52}$Cr BEC with $a_{\mathrm{dd}}=15a_0$
for $a=100a_0$ and different $N$ in Figs. \ref{fig2} (a) for $S(0)=0.4$ and (b) for  $S(0)=0.8$.  First we consider the results for $S(0)=0.4$. For a very small $N$ ($=10$) we have Josephson oscillation with small frequency. This value of $N$ is below the critical value $N_c$  for self trapping as given by Eq. (\ref{crit}). For a larger $N$ ($=100$) this critical value is achieved $(N>N_c)$
and self trapping is encountered with $\langle S(t) \rangle\approx 0.36$. {Finally, for a much larger $N$ ($=5000$), above an upper critical limit $N_u$ ($N>N_u$),
the nonlinear interactions are very large, while the Gaussian barrier in the trapping potential (\ref{sw}) can be neglected and the double well essentially reduces to a single well and the dipolar BEC executes free
sinusoidal  oscillation.}  However, for the  number of atoms  $N$ slightly greater than the upper limit $N_u$ ($N_u\lesssim N$), the oscillation is irregular and not regular. A smooth sinusoidal oscillation appears for a much larger value of $N$ ($N>>N_u$).   
  A similar panorama takes place for $S(0)=0.8$ as shown in Fig. \ref{fig2} (b). As $S(0)$ is larger in this case, the limiting value of nonlinearity for self trapping is smaller in this case and we already have self trapping for $N=10$, whereas for this $N$ for $S(0)=0.4$ we have Josephson oscillation, viz. Fig. \ref{fig2} (a).  For a large enough value of $N$ there should be free sinusoidal oscillation of the system, which has not appeared for $N=10000$.

\begin{figure}[!t]
\begin{center}\includegraphics[width=\linewidth]{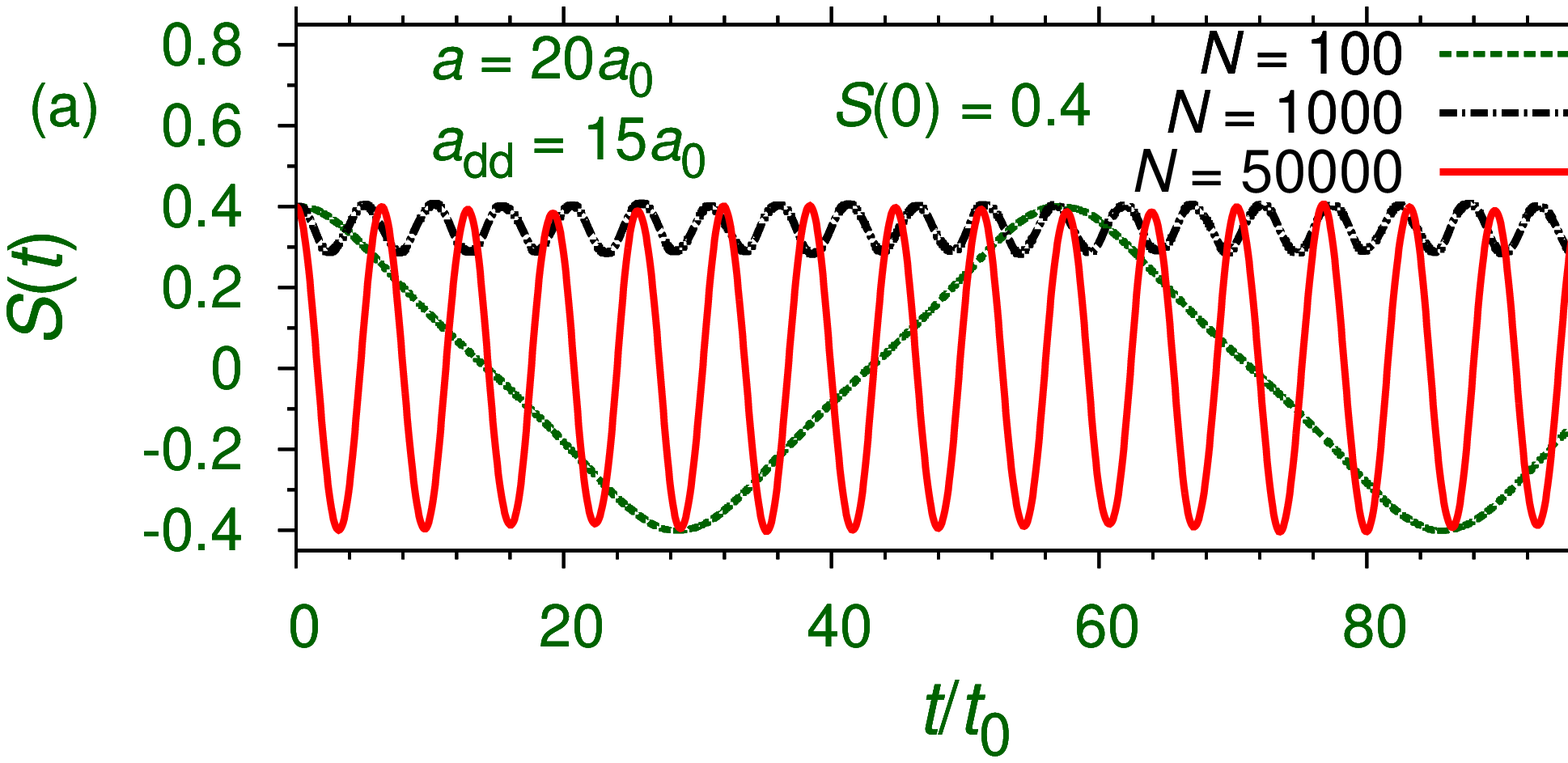}
\includegraphics[width=\linewidth]{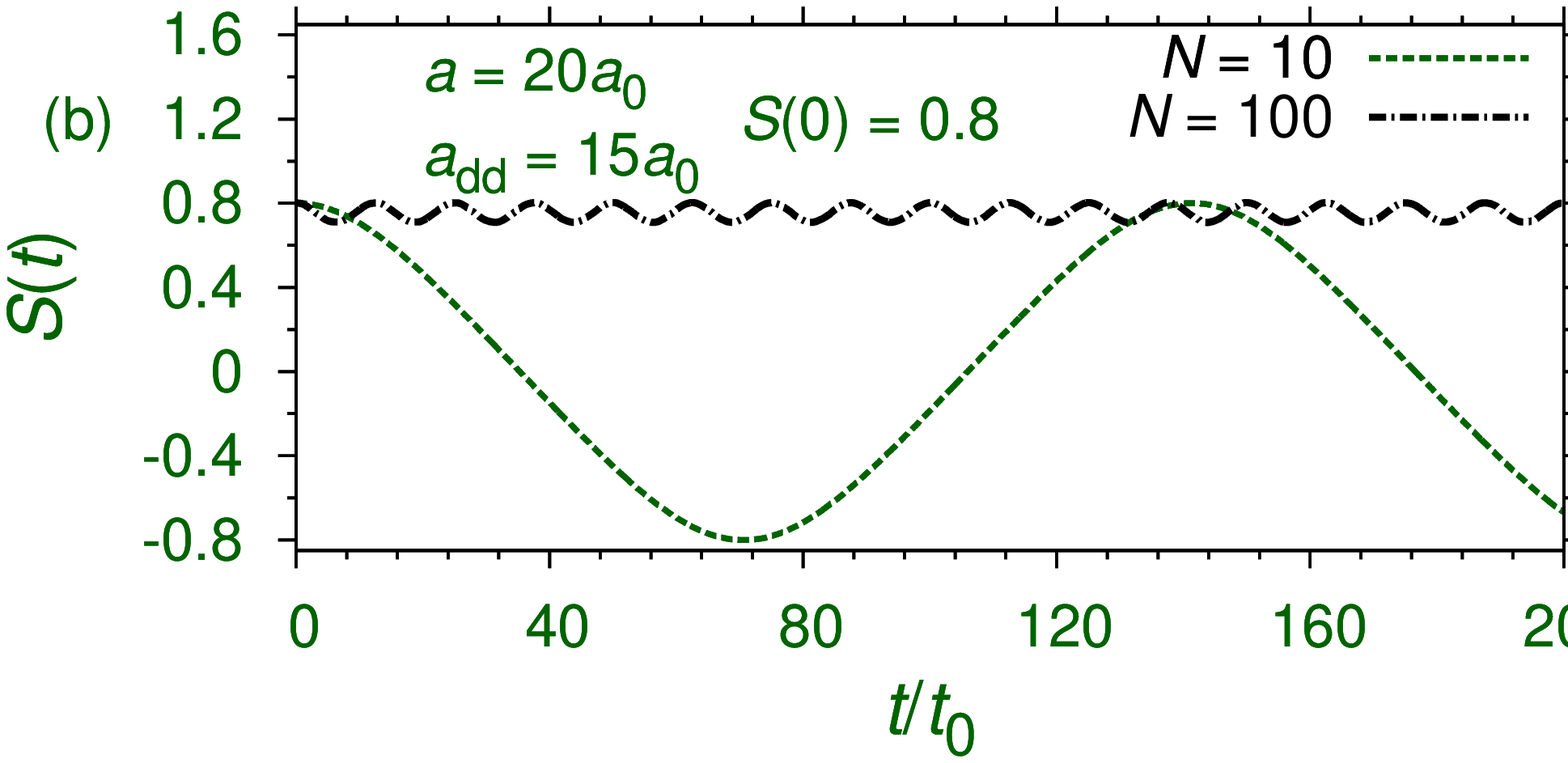}

\caption{ (Color online)  Population imbalance $S(t)$ versus time $t/t_0$ 
for  $a=20a_0, a_{\mathrm {dd}}=15a_0$ and for  (a)
 $S(0)=0.4$ and (b)  $S(0)=0.8$ and  
 for different $N$ using the 1D model (\ref{gp1d}).
  }\label{fig3}
\end{center}
\end{figure}

Similar physics appears for $a=20a_0$ in a cigar-shaped dipolar $^{52}$Cr BEC  as presented in 
Figs.  \ref{fig3} (a) for $S(0)=0.4$ and (b)  for $S(0)=0.8$.  However, the lower critical limit  $N=N_c$ for  self trapping  for $a=20a_0$ 
is much larger compared to the dynamics presented in Figs. \ref{fig2} for $a=100a_0$. The net repulsive interaction with a smaller scattering length $a$ in this case is much smaller for a fixed $N$ compared to the net repulsive interaction in Fig. \ref{fig2} with a larger $a$. Consequently, the desired repulsive nonlinearity for self trapping is achieved for a larger $N$ in Figs. \ref{fig3}. Otherwise, the dynamics presented in Figs. \ref{fig3} is consistent with the theoretical expectation.  With the increase of $N$, the dynamics passes from the oscillatory regime to self trapping and then back to the oscillatory regime again.

\begin{figure}[!t]
\begin{center}\includegraphics[width=\linewidth]{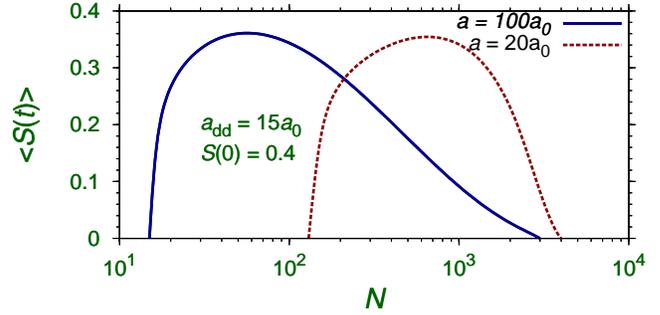}
\caption{ (Color online) Time-averaged population imbalanced $\langle S(t)\rangle $ versus the total number of atoms $N$ for $a=100a_0$ and $20a_0$ for an initial population imbalance $S(0)=0.4$ using the 1D model (\ref{gp1d}).  
  }\label{fig4}
\end{center}
\end{figure}

For  experimental interest, the  phenomenon of self trapping and Josephson oscillation is well illustrated through an exposition of time-averaged population imbalance $\langle S(t) \rangle $ versus the total number of atoms $N$ for a given set of dipolar and contact interactions. In Fig. \ref{fig4}  we plot  $\langle S(t) \rangle $
versus $N$  for $ a=100a_0 $ and  $ a=20a_0, $ with the parameters  $S(0)=0.4$ and $a_{\mathrm{dd}}=15a_0$.   For $ a=100a_0 $
self trapping appears for $N>N_c\approx 16$ whereas  for $ a=20a_0 $ self trapping appears for $N>N_c\approx 145$. In the first case the net atomic interaction is more repulsive because of the larger value of scattering length $a$. Consequently, the limiting nonlinearity for self trapping is achieved for a smaller number of atoms. In the second case the net atomic interaction is weakly repulsive because of a smaller value of scattering length and as $a\approx a_{\mathrm{dd}}.$
Consequently, the limiting nonlinearity for self trapping is achieved
 for a larger number of atoms. After the onset of self trapping with $N$ past the critical number of atoms, $\langle S(t)\rangle$ first increases and approximates $S(0)$. With further increase of $N$,     $\langle S(t)\rangle$ eventually becomes zero while there cannot be any self trapping. The absence of self trapping for large $N>N_u$ takes place when the small Gaussian barrier in the trapping potential becomes very small compared to the nonlinear terms and its effect can be neglected in the GP equation. Consequently, the double well effectively reduces to a single well, where there cannot be any self trapping. 

In this investigation, we took the angle $\varphi$ between the polarization direction $z$  and the  double-well orientation to be zero. The dipolar interaction in this configuration is attractive.  If the double-well orientation is taken along the $x$ direction with $\varphi = \pi/2$, the dipolar interaction will be repulsive. 
Consequently, for dynamics along the  double-well orientation direction $x$, self trapping should be possible for the initial population imbalance $S(0)$ above a critical value and for all values of the scattering length $a$ $(>0)$. 
It would be interesting to study this nontrivial self trapping in a fully anisotropic environment.

\section{Summary and Conclusion}

\label{IV}

We studied the dynamical self trapping and Josephson oscillation 
of a repulsive cigar-shaped dipolar $^{52}$Cr BEC trapped in an axially-symmetric  double-well potential aligned along the polarization direction. The dipolar BEC was subject to a strong radial and weak axial confinement and we use an effective 1D mean-field model appropriate for the description of its dynamics. Although, most of the results presented here were obtained  using the 1D model, some of these were also confirmed using the full 3D model from which the effective 1D model was obtained.  This assures that the results obtained with the 1D model will not be so peculiar as to have no general validity.    Two values of the scattering length were considered: 
$a=100a_0$ and $20a_0$  The former corresponds to a rounded-up value of the experimental \cite{pfau,4} scattering length and the later chosen 
to keep the dipolar BEC weakly attractive.

The two-mode model originally proposed for a description of self trapping of a repulsive cigar-shaped nondipolar BEC in a double-well 
potential was extended to include an  additional dipolar interaction. 
This modified two-mode description could explain the essential features of Josephson oscillation and self trapping of the cigar-shaped dipolar BEC, which are the following.  The phenomenon of Josephson oscillation and self trapping is very sensitive to the total number of atoms $N$ and the initial population imbalance $S_0$. Self trapping takes place for $S_0$ greater than a critical value $S_c$ and $N$ between two limiting values $N_u>N>N_c$. This study was performed with realistic values of trapping parameters for a dipolar $^{52}$Cr
BEC with realistic values of dipolar and contact interactions so that the results can be compared with possible future experiments.


\acknowledgments
We thank FAPESP  and  CNPq (Brazil)  for partial  support.


\begin{thebibliography}{99}
\bibitem{footnote} Similar, but not identical, two-mode equations 
were obtained in Ref. \cite{xiong}, viz.  Eqs. (8) in that reference. 
The reason for the difference is not clear.
\end{thebibliography}
\end{document}